\documentclass[11pt]{article}
\usepackage{fullpage}
\usepackage{graphicx}
\PassOptionsToPackage{hyphens}{url}\usepackage{hyperref}

\title{Flipping the Perspective in Contact Tracing}

\author{Po-Shen Loh\thanks{Department of Mathematical Sciences, Carnegie
  Mellon University; and Expii, Inc. Email: \texttt{po@novid.org}. Research
supported in part by the Tansy Charitable Foundation.}}

\begin{document}
\maketitle

\begin{abstract}
  We introduce a fundamentally different paradigm for contact tracing: for
  each positive case, do not only ask direct contacts to quarantine;
  instead, tell everyone how many relationships away the disease just
  struck (so, ``2'' is a close physical contact of a close physical
  contact). This new approach, which has already been deployed in a
  publicly downloadable app, brings a new tool to bear on pandemic control,
  powered by network theory. Like a weather satellite providing early
  warning of incoming hurricanes, it empowers individuals to see
  transmission approaching from far away, and incites behavior change to
  directly avoid exposure.  This flipped perspective engages natural
  self-interested instincts of self-preservation, reducing reliance on
  altruism, and the resulting caution reduces pandemic spread in the social
  vicinity of each infection. Consequently, our new system solves the
  behavior coordination problem which has hampered many other app-based
  interventions to date.  We also provide a heuristic mathematical analysis
  that shows how our system already achieves critical mass from the user
  perspective at very low adoption thresholds (likely below 10\% in some
  common types of communities as indicated empirically in the first
  practical deployment); after that point, the design of our system
  naturally accelerates further adoption, while also alerting even
  non-users of the app. This article seeks to lay the theoretical
  foundation for our approach, and to open the area for further research
  along many dimensions.
\end{abstract}

\section{Introduction}

The COVID-19 pandemic has severely impacted daily life across much of the
world. Due to extensive public discussion and large-scale implementation,
the paradigm of traditional contact tracing has reached conversational
familiarity: test for positive individuals, and then trace their proximal
contacts, so as to isolate them from potentially infecting the rest of the
community \cite{bib:cdc-contact-tracing}.

Relatively early in the pandemic, many projects sprang up to develop apps
to digitize and accelerate this process (see, e.g., the book
\cite{bib:book-kahn}, the media reports \cite{bib:techcrunch-15-apps,
bib:wsj-apps-not-ready}, or the list \cite{bib:wiki-covid-19-apps}).
Although these apps were very numerous, the entire class of apps sought to
deliver essentially the same primary value to the user: after the user
became directly exposed to a COVID-positive individual, the app would
notify them to take a variety of actions designed to protect the rest of
society from the app user (who might now be infected). These apps achieved
this goal through a variety of sensing methods, ranging from scanning QR
codes \cite{bib:nhs-qr}, to GPS \cite{bib:marketwatch-gps}, Wi-Fi
\cite{bib:tracefi}, Bluetooth \cite{bib:trace-together}, and ultrasound
\cite{bib:loh-novid-accuracy, bib:cmu-novid-accuracy}. A lively debate
emerged over whether the infrastructure for delivering this user experience
should be centralized or decentralized
\cite{bib:centralized-vs-decentralized}, ultimately tilting in favor of
decentralized architectures, consistent with the principle of seeking the
least privacy-invasive approach for delivering the post-exposure
notification user experience.

Although many prominent governments strongly preferred a centralized
approach, the benefits of centralization were primarily explained from the
perspective of the central government being able to make better decisions
for the common good. However, there was substantial pushback against the
collection of a centralized interaction network by a government or large
corporation which already inherently had access to personal information
(e.g., in other databases within the government or other business units
under the same organizational umbrella), as that would compromise
anonymity.  The centralization vs.\ decentralization debate was ultimately
quelled by a historic collaboration between Apple \cite{bib:gaen-apple} and
Google \cite{bib:gaen-google}, whereby they integrated a decentralized
framework into their mobile operating systems, and strictly regulated
access to it.  Specifically, their system provided the only way for iPhones
to reliably detect nearby iPhones via Bluetooth under all circumstances. At
the time, there were significant complaints from the national governments
of France, Germany, and the United Kingdom about decentralization
\cite{bib:france-germany-uk}, but the system remained.

Of the three, Germany was the first to deploy the decentralized approach,
but after investing 25 million euros to develop and promote the resulting
app, the resulting impact is in question.  One report
\cite{bib:germany-gaen-disappoint} noted the weakened power of
decentralization and the research documenting the inaccuracy of Bluetooth
\cite{bib:tcd-bluetooth, bib:tcd-bus, bib:tcd-tram}. Unfortunately, the
pandemic continues to spread in many regions of the world, with the United
States surpassing 9 million cases on October 29 \cite{bib:usa-9-million}.

To estimate the effect of the now-standard exposure notification technique,
a recent preprint \cite{bib:openabm-google} by researchers from Google and
Oxford reported that ``in a model in which 15\% of the population
participated, we found that digital exposure notification systems could
reduce infections and deaths by approximately 8\% and 6\%,'' under the
assumption that when someone anonymously receives an exposure notification,
which is tuned so that ``80\% of all `too close for too long' interactions
are captured between users that have the app,'' they voluntarily are ``90\%
likely to begin a quarantine until 14 days from initial exposure with a 2\%
drop out rate each day for noncompliance.''

Another recent preprint \cite{bib:covidwatch}, coauthored by people
affiliated with a leading app using the Google-Apple system, contextualized
the probability of infection associated with exposure notifications,
indicating that 14-day quarantines could be requested for individuals when
their infection risk exceeded a threshold around 1\%, indicating that such
a figure was ``broadly compatible with the attack rate reported in Taiwan
(1.0\%, 95\% CI: 0.6--1.6\%) for those interacting with infected
individuals in the first 5 days of symptom onset \cite{bib:taiwan}, which
is similar to the 1.9\% attack rate (95\% CI 1.8\%--2.0\%) reported in
South Korea \cite{bib:sk}.'' In light of these figures, it is likely that
in order to capture 80\% of all interactions that result in infection, the
probability of being infected upon receiving an anonymous exposure
notification would be under 10\%, due to limitations from anonymization and
technology accuracy.

It is realistic in the non-anonymous manual contact tracing setting to
expect 90\% initial quarantine compliance, because each affected contact is
directly contacted by the authorities who know their identity (and who may
be able to enforce policies), and the manual contact tracer is likely to
have taken other factors into account (e.g., indoor/outdoor, usage of
personal protective equipment) to significantly reduce false positives. On
the other hand, it is less clear that there would be 90\% compliance with
non-enforceable anonymous exposure notifications corresponding to
under-10\% infection probability. Indeed, if the voluntarily compliance
rate upon receipt of low-infection-probability exposure notifications were
sharply lower than 90\%, then even if there were 100\% app adoption, the
entire intervention would have minimal impact.

These observations indicate that it is very important to explore other
interventions. They also indicate that since human behavior plays a major
factor in driving efficacy, it is valuable to consider the behavioral
science perspective, as compliance with post-exposure notifications
requires altruism. There has been substantial work in the behavioral
science domain discussing COVID-19 in particular (see, e.g., the surveys
\cite{bib:behavior-survey-nature, bib:behavior-survey-jbpa}), and it would
be beneficial to design a system which naturally aligns with human
behavior.

In this article, we introduce a fundamentally different approach to the
general problem space of contact tracing, which realigns incentives with
self-interest, to boost both the initial app adoption and the response to
signals from the app. It can seamlessly coexist with post-exposure
notification (manual and/or digital), and we recommend both interventions
to be deployed in parallel. We summarize our system in the next section,
and provide a detailed specification for a sample full implementation in
Section \ref{sec:system-spec}. Then, Section \ref{sec:heuristic} presents
some empirical observations from actual deployments, together with a
heuristic mathematical analysis of how the natural incentive alignment in
our approach is likely to very significantly boost effectiveness. We
address privacy and security issues in Section \ref{sec:privacy-security}.
The purpose of this article is to introduce our new approach, demonstrate
its substantial theoretical advantages, and open the large area of
investigation around it, which may be of interest to researchers from
disciplines ranging from public health to behavioral science to
mathematical modeling.

\section{New approach}

We use the physical interaction (contact) network from a different
perspective.  Abstractly, that network represents the potential disease
transmission network, and may be enriched and extended by attaching a
variety of times, strengths, and other properties to describe the natures
of interactions and individuals.  In the context of this network, the
standard contact tracing paradigm is: for each newly positive node $v$,
send post-exposure notifications to the neighboring nodes $N(v)$ which are
directly connected to $v$ in the network by sufficiently-strong connections
that occurred at times overlapping with $v$'s contagious period. The
existing intervention achieves its main impact on disease spread by
quarantining (and testing) the nodes in $N(v)$, with the hope of cutting
off a significant fraction of the spread of the disease.

\begin{figure}[h]
  \begin{center}
    \includegraphics[width=0.4\linewidth]{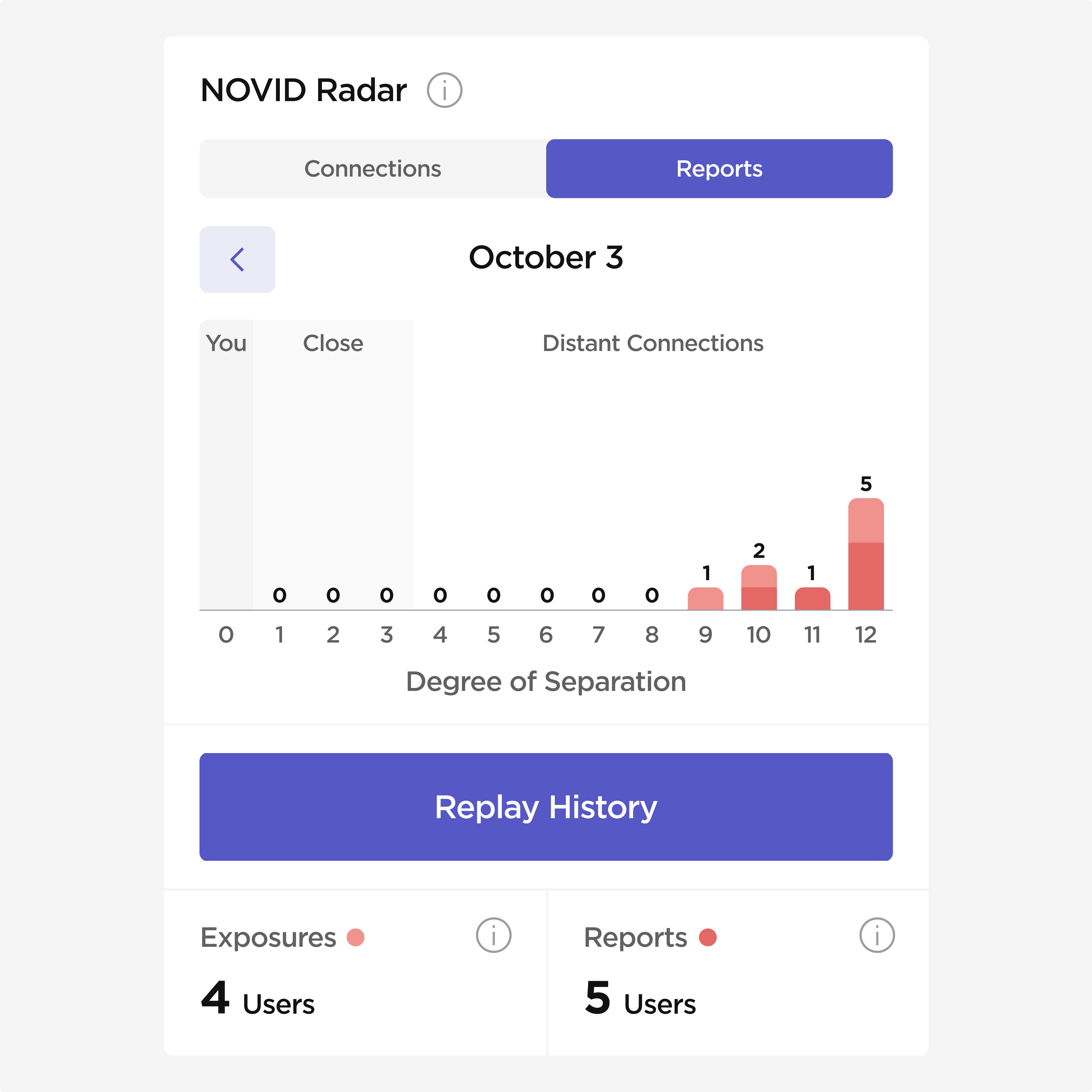}
    \hspace{2mm}
    \includegraphics[width=0.4\linewidth]{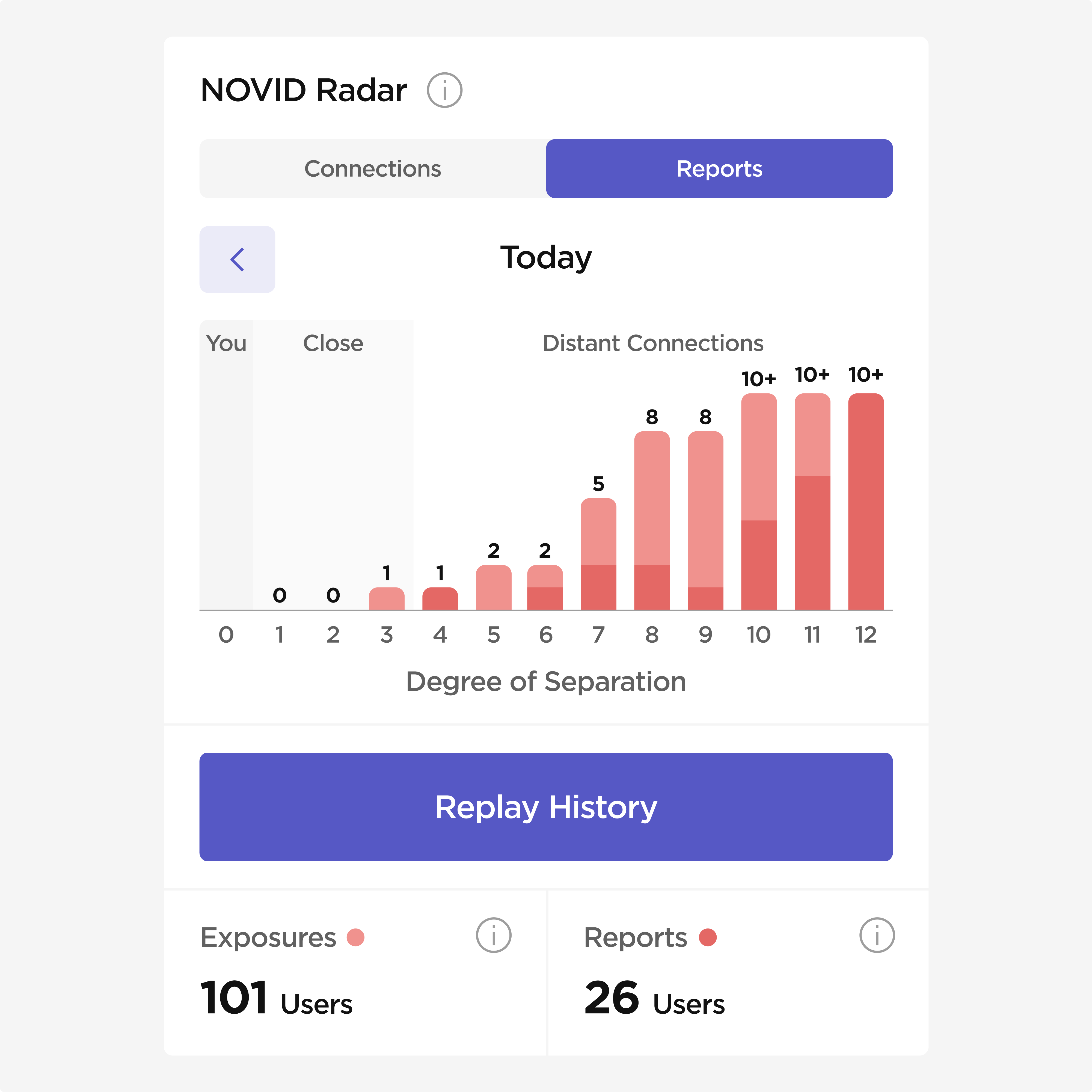}
  \end{center}
  \caption{\footnotesize \emph{Two representative animation frames,
    indicating the approach of the infection over a multi-day period. The
    height of each bar represents the number of users who have tested
    positive (dark red) or been confirmed as contacts of positive cases
    (light red), and the horizontal axis is the network distance between
    the positive case and you, where the network is constructed based on
    the last 14 days.  Positive cases disappear from the chart after a
    specified period.}}
  \label{fig:network-chart}
\end{figure}

Our approach flips the perspective from that of the central planner seeking
to reduce spread, to that of the individual seeking to avoid infection.
For each newly positive node $v$ in the network, we notify all other nodes
connected by paths of network distance\footnote{The \emph{network distance}
between two nodes in a network is the minimum number of connections that
need to be traversed to go from one node to the other.  For example, two
nodes that are directly connected are at network distance 1; and, two nodes
that are not directly connected to each other, but are both directly
connected to the same other node, are at network distance 2.} up to some
value (e.g., 12 in our implementation as of November 2020) of the numerical
network distance that they are from $v$, but not the direct identity of
$v$. The network used is based on all recent physical interactions across
the entire user population, as detailed in Section
\ref{sec:network-construct}. By continually providing this network distance
information and animating it over time like a weather radar map, every
individual user can visualize the infection approaching or receding in
their own network. This is illustrated in Figure \ref{fig:network-chart}.

Our ``pre-exposure notification'' system provides lead time as compared to
traditional exposure notification (which only notifies nodes at network
distance 1), akin to how a hurricane satellite video empowers people to
take precaution before the storm is directly overhead.  By seeing
transmission approaching from far off, the individual has time to protect
themself. Since there is lead time, the user's context is fundamentally
different. Instead of being asked to take actions that protect others from
themselves (as would be necessary if they already had direct contact with a
confirmed-infected individual), they have the opportunity to protect
themselves from others. This important asymmetry enables our intervention
to achieve its main impact on disease spread by dynamically engaging
natural self-interested instincts of self-preservation, as people near each
infection hotspot take actions to avoid infection, such as using stronger
protective equipment, practicing more vigilant distancing, etc.

By focusing on the individual's perspective, our approach aims to deliver
actionable information to individuals in a way where they clearly perceive
direct value. We are definitely not the only ones to propose tools that let
individuals visualize their risk (e.g., \cite{bib:forbes-klick}) or
visualize the spread in geographical regions (e.g.,
\cite{bib:google-maps}). However, it is significantly more actionable and
relevant to learn of a new case at network distance 3 from you (maybe your
housemate's coworker's husband Bob), rather than to learn that there were 5
new cases in your postal code. Indeed, Bob might live on the other side of
town.

Conveniently, the evolution of positive cases is relatively easy to plot as
a bar chart, to animate against network distance on a smartphone display.
Figure \ref{fig:network-chart} displays two animation frames of this simple
interface, which even to a non-technical audience communicates the alarming
approach of infection from far away. This animated visual provides a
naturally easy-to-use tool for any individual to avoid infection,
regardless of their technical background.  There is no abstract risk score
to interpret, nor an infection probability to estimate after direct
contact. That is because our objective is not to quarantine the user to
prevent them from infecting others, but rather to inspire the user to
increase caution to avoid getting infected. All they need to do is to watch
the bars animating day-by-day, so they can take extra precaution when the
bars seem to be on track to reaching the user. Such is the value of
notifying all users of their numerical network distance from each new
positive case, instead of only notifying the users at network distance 1
that they have already been exposed.

Our approach thus effectively becomes an automatic way to dynamically
modulate social distancing in sub-regions of a community, as people close
to hotspots (in network distance) see more red on their charts. Network
distance is the correct metric, and is more appropriate than geographic
distance because it accounts for people traveling to workplaces in
different parts of town. If people just took greater precaution (e.g., more
vigilant use of personal protective equipment, more intense distancing,
etc., but not even a rigorous quarantine) whenever the virus struck within
close network distance (e.g., 3) of them according to their chart, then
this temporary behavior change would tend to reduce the person-to-person
transmission by a significant factor in the vicinity of every infection
cluster. We hypothesize that this would already have a nontrivial impact on
the infection's basic reproduction number ($R_0$) in the population.

\section{System specification}
\label{sec:system-spec}

Our framework is currently deployed as an app, which has been publicly
available for free download from the official Apple and Google app stores
since May 2020 \cite{bib:novid-app}. To provide the reader with a concrete
example of an implementation, we sketch that particular app's construction
in this section. That app actually also provides post-exposure
notifications as well, but this article will concentrate on the
pre-exposure notification component. The purpose of this article is not to
focus on one particular implementation of our approach, but this section
seeks to include sufficient detail to demonstrate that our approach is
practically implementable in today's environment. We describe the different
sensing mechanisms because their practical idiosyncrasies profoundly impact
the feasibility of our framework. It is interesting that this system only
recently became possible to implement at wide scale, thanks to the
proliferation of smartphones possessing short-range communication
capabilities.

\subsection{Pseudonymous network construction}
\label{sec:network-construct}

Strictly speaking, our system is pseudonymous rather than anonymous, in the
sense that each user has a persistent Version 4 Universally Unique
Identifier (UUID), which is randomly generated at install time and never
revealed to the user, nor associated with personally identifiable
information. Importantly, it is impossible to recover any hardware
identifiers or phone numbers corresponding to the user's smartphone from
their UUID.  The app then periodically scans its vicinity to identify
proximal app-running devices using Bluetooth Low Energy, ultrasound, and
Wi-Fi.

\begin{description}
  \item[Bluetooth Low Energy (BLE).] Similarly to most other apps in this
    space, this app wakes up every few minutes to scan the vicinity for
    several seconds, searching for other BLE devices running the same app.
    If it finds any, the devices communicate over BLE to exchange temporary
    random identifiers (not their UUID's), and they each independently send
    our central server their UUID's, the temporary identifiers they sent
    and received, the current time, and the received signal strength
    indicator (RSSI). Due to iOS limitations, a backgrounded iOS
    app\footnote{Generally speaking, an app is running in the foreground
    when it is on screen; it is running in the background when it is not on
  screen or the phone screen is off.} cannot detect other backgrounded iOS
  apps when scanning, but it can be detected by foregrounded iOS apps and
  Android apps in both backgrounded and foregrounded states, whereby it
  will respond to scans from those devices. Android devices have no
  significant limitations.

  \item[Ultrasound.] During the BLE communication, if neither device is a
    backgrounded iOS app, then the devices use a near-ultrasound
    communication protocol in the 18.5--19.5 KHz range, to estimate their
    relative distance with a significantly higher level of accuracy than
    can be deducted from BLE RSSI (see, e.g.,
    \cite{bib:cmu-novid-accuracy}). Since ultrasound does not penetrate
    walls, this additional sensing technique verifies whether the devices
    are in the same airspace. Custom sonic waveforms are used to optimize
    robustness in noisy environments with obstacles.  Importantly, all
    Fourier Analytic signal processing is handled on-device, and no audio
    recordings are stored on the device or transmitted to a server for
    processing. Only the estimated distance is transmitted to the server,
    augmenting the BLE data record mentioned above. The implementation of
    this Ultrasound capability is non-trivial, and falls beyond the scope
    of this article. A future technical publication will address it in
    greater detail.

  \item[Wi-Fi.] Every few minutes, wake-up signals are sent from our own
    remote server to phones, as well as via Bluetooth from Android phones
    or foregrounded iOS apps. Upon receiving a signal, the app checks if it
    is connected to a Wi-Fi access point.  If it is, it sends a hashed
    version of the specific access point's fingerprint (technically, its
    BSSID) to a separate server without including the user's UUID. That
    server returns a temporary and randomly generated identifier for that
    BSSID which is only stable for a fraction of an hour. All associations
    between BSSID's and temporary identifiers are rapidly expired and
    expunged. The app works only with that temporary Wi-Fi identifier, and
    sends it to our main server together with the time at which this check
    occurred. This provides the ability to detect whether two devices were
    connected to the same Wi-Fi access point at around the same time, while
    making it difficult to use the interaction database to reverse-engineer
    the original BSSID's from the stored Wi-Fi information.

  \item[No GPS.] Very importantly, GPS is never used, because a device's
    GPS coordinates constitute personally identifiable information.

  \item[No constant Internet connectivity required.] Among some
    populations in the world, many people only have intermittent Internet
    connectivity.  Those populations tend to correlate with heavier Android
    use, at which point our BLE/ultrasound sensing fully operates even
    without the Internet. Whenever the device reaches Internet
    connectivity, detection records created since the last instance of
    connectivity are uploaded for server processing, and the latest
    notification information is downloaded. Even iPhone users without
    cellular data can use a ``Standby Mode'' which we created to keep the
    app in the foreground to use BLE/ultrasound while dimming the screen to
    conserve battery.

\end{description}

The Wi-Fi sensing component is the unique element that enables our system
to operate on iPhones even when the app is in the background. Its
resolution is insufficient for the purpose of post-exposure notification,
and so it was not an option for apps using that existing paradigm.
Background iOS operation was a fatal issue for other apps. Unfortunately,
the United Kingdom's initial attempt to build its own Bluetooth app outside
of the specially-regulated Google-Apple Bluetooth system only detected 4\%
of iPhones \cite{bib:isle-wight}, understandably leading them to abandon
the project.

The power of our alternative paradigm is that it makes Wi-Fi extremely
useful, because our method only needs to construct an approximate physical
interaction network. Our method of impact is to provide anonymized insights
which confer the same degree of risk as ``your housemate's coworker's
husband Bob became positive,'' but without identifying that the
relationships in question are those specific co-living or co-working ones,
instead stating that someone at network distance 3 became positive.  In
many situations, these types of strong relationships (co-living and
co-working) are reliably captured by Wi-Fi access point overlaps of several
hours. These are the types of relationships that form much of the backbone
of the network along which the infection spreads.  Indeed, one of the
most-referenced COVID-19 agent-based modeling software packages,
OpenABM-Covid19 \cite{bib:openabm}, models the infection network by
overlaying ``household networks'' and ``occupation networks,'' together
with some random interactions. Our Wi-Fi system effectively captures a
large fraction of the household and occupation networks, supplemented by
Bluetooth and ultrasound to provide additional accuracy and valuable
redundancy. To capture the random interactions, we tune our Bluetooth and
ultrasound parameters to be fairly generous, picking up contacts if they
have been within around 10 meters for at least 15 minutes.

In summary, the app's 3-sensor system provides multiple overlapping
technologies with different strengths and weaknesses. We use 14-day windows
of data to construct approximations of the physical interaction network,
with 14 days chosen in order to capture two weekends for redundancy,
because people often have different interaction patterns on weekends. This
suffices to construct an approximation of the network which serves the
purpose of our new paradigm, and unlocks its potential.

\subsection{Anonymous positive case labeling}
\label{sec:positive-token}

In order to reliably label some nodes as positive, we use a one-time token
system which has a very important improvement over other apps' systems that
vastly amplifies our approach's power: in our system, tokens can be
entered\footnote{In some alternative implementations of our system, users
may be able to pre-authorize a trusted authority to enter signals on their
behalf, by submitting a token of their own.} not only by confirmed positive
cases, but also by confirmed contacts of positive cases, as confirmed by a
contact tracer. The latter type of signal is not useful for the traditional
quarantine-driven paradigm, because the infection risk upon being a contact
of a contact of a positive case is far too low to recommend quarantine. It
is unique to our alternative distance-based paradigm.

Specifically, a trusted authority (e.g., government department of public
health, university health center, etc.) is able to securely generate tokens
from a separate system we operate, to which they input no personal data to
generate each token. They then distribute the tokens to individuals in
their jurisdiction who they have confirmed to be positive.  When a user
enters a recognized token into their app, together with the date symptoms
started, we mark their UUID as positive.  Our databases do not store the
association between which UUID used which token. For evaluation purposes,
it is also possible to submit a positive report in the general app without
a one-time token, but official community deployments typically disable
unauthenticated reporting.

In addition, the trusted authority can generate a different type of tokens
to distribute to contacts of confirmed positive cases during their
(possibly manual) contact tracing process.\footnote{In our deployments, we
  do not store any linkage between the tokens that correspond to each
  individual case, to preserve privacy. In settings where the resulting
  lack of privacy is acceptable, it is possible to store such linkages.
  Then, the system could even estimate the distance from Person $A$ to a
  positive case for which only contact tokens were submitted, by adding 1
  to the minimum of all network distances from Person $A$ to people who
  submitted tokens linked to that positive case. Other natural variations
would also be possible.} Users can enter these confirmed contact tokens
into their app, so that other users can see how many relationships away
they were from people who were identified as confirmed contacts of some
positive case during the contact-tracing process. This is still useful in
our paradigm because if a user sees that there is such a person $d$
relationships away, then they are less than or equal to $d+1$ relationships
away from a confirmed positive case. Yet this significantly amplifies the
probability that each positive case generates some signal in the system.
Indeed, if a positive case had 10 contact-traced contacts, then even if
each of the 11 tokens independently only had 20\% chance of being entered,
there is now over 90\% chance that at least one of the 11 people enters
their tokens.

\subsection{Chart construction}
\label{sec:chart-construct}

To minimize privacy concerns, we only send new signals to other users at
the time each new positive test token is entered into the mobile app. In
particular, we do not continue to send new signals that would reveal
whether the positive person chooses to continue walking around in public.
In order to make the positive signals more visible on the animated chart
from each particular user's perspective (Figure \ref{fig:network-chart}),
when our server discovers that there is a newly positive person, that
person contributes $+1$ to the chart at the network distance to the user
based on the network constructed from the last 14 days of population-wide
interaction data.  Then, the $+1$ contribution from this positive case
remains on the user's chart at this fixed distance until 10 days after the
reported symptom start date (duration constantly evaluated based upon
latest guidance), similarly to how an old-fashioned radar screen displays
glowing blips that take some time to fade away. Confirmed contacts of
positive cases (as entered via the contact codes in the previous section)
are overlaid on this chart in a lighter color.

\section{Heuristic analysis and empirical observations}
\label{sec:heuristic}

An essential question for all digital interventions is to determine what
adoption level is required for impact. Other approaches which primarily
protect other people from the app user, or more generally do not provide a
direct benefit to the user, tend to have greater difficulty intrinsically
motivating adoption.  In contrast, our approach takes inspiration from apps
that enjoy organic growth, by delivering the direct value of helping the
user proactively protect themself.  Therefore, our approach raises two
adoption-related questions.  First, since our approach leverages network
effects, what is the critical adoption threshold beyond which users
conclude that a critical mass of people in their region have adopted this
solution, to facilitate organic growth?  Second, at what (possibly higher)
threshold does the app deliver enough impact to make a meaningful impact on
infection spread? It is worth noting that the existence of the first
question is a major benefit, because it provides a roadmap to high adoption
from a low initial threshold. It turns out that the physical interaction
network very significantly amplifies the power of our approach.

\subsection{Low threshold for critical mass}

In this subsection, we include some empirical observations from the first
community deployment of an earlier version of the app based on our
approach, at Georgia Tech \cite{bib:gatech-deploy}, where the community was
encouraged to participate on a voluntary basis. It is important to note
that although the particular app version provided significant function, it
did not yet have the Wi-Fi-based capability to operate fully in the
background on iOS devices, and so required more user intervention in order
to operate at all times. No confidential data collected from any deployment
is referenced in this section.

\begin{figure}[h]
  \begin{center}
    \includegraphics[trim={10 285 10 380},clip,width=0.3\linewidth]{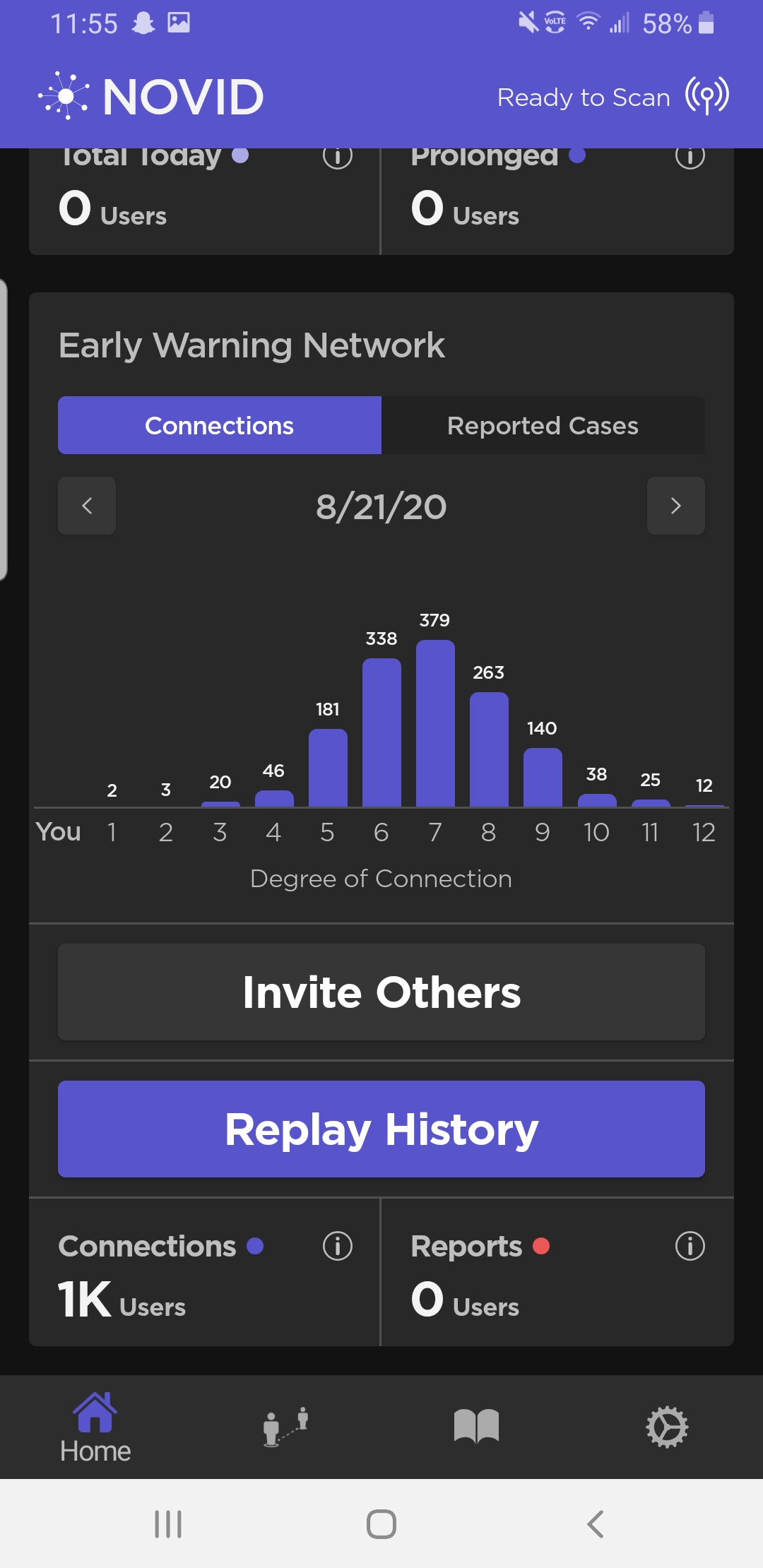}
  \end{center}
  \caption{\footnotesize \emph{This chart, publicly shared by a Reddit user
    \cite{bib:reddit-gatech}, plots the number of other users at each
    network distance from the user in an active deployment. Even though the
    particular user only had 2 direct connections, there is an exponential
  growth in the number of users until the perimeter of the connected
cluster is reached.}}
  \label{fig:network-chart-blue}
\end{figure}

User feedback during the rollout period pointed to the value our
network-based approach provided in indicating personally-relevant critical
mass. As reported by an independent student newspaper at the unaffiliated
University of Georgia \cite{bib:uga-gatech}, when the app was at around a
10\% level of adoption at Georgia Tech, one interviewed user noted they had
nearly 2,000 connections at various distances from them in particular.  The
app displayed this to them in a chart analogous to Figure
\ref{fig:network-chart-blue}, which is itself a screenshot shared in a
Reddit post \cite{bib:reddit-gatech} by a different student who said: ``I
checked in yesterday, and I seem to have my room mates on there, and the
charts are unbelievable. I can't believe it linked our little enclave to
much of campus residents.'' This critical mass effect also inspired
installation by users outside of the Georgia Tech campus community, as a
parent independently communicated that he had installed the app, noting the
1,000+ connection counts that he observed on his app due to connecting with
the student during regular home visits. The fact that people highlighted
these individually relevant connection counts indicates that our
network-based approach already helps people perceive critical mass at low
adoption rates.

There is heuristic mathematical support for this observation that in dense
environments such as universities and schools, a low adoption threshold
suffices to enable users to have many other users connected to them in the
system. This corresponds to the property that our collected network
contains large connected clusters, where within each large cluster, every
pair of nodes is connected by at least one path (possibly via other nodes).
This is not the same as effectiveness, but it does answer the first
question above of helping people in a region conclude that critical mass
has been reached for the app user network to be connecting them to many
participants. Our system (as of November 2020) builds its network by looking
over 14-day windows, and identifying app users who have been detected
within about 10 meters\footnote{Since our alternative approach is not
  focused on sending exposure notifications to trigger quarantine, it is
more useful to have a wider radius for establishing connections that
indicate relationships, as opposed to the commonly-used value of 2 meters
that is used for transmission.} for at least 15 minutes, or concurrently on
the same Wi-Fi access point for at least 3 hours. Charts with large numbers
like Figure \ref{fig:network-chart-blue} start to appear when a typical
user is connected to at least 2 other app users in this network, and each
of them is connected to at least 2 other users, and so on. In some sense,
we are building a viral app which spreads along a similar network to the
virus itself, and this corresponds to the app having its own $R_0 \geq 2$.
This type of phenomenon roughly emerges when the adoption rate in a region
of the actual network (e.g., a university or school) exceeds about
$\frac{3}{d}$, where $d$ estimates a typical person's number of contacts as
defined with the generous parameters above over a 14-day period. For
schools and universities with in-person classes or students living in
residence halls, $d$ may be over 30; then, this critical threshold is below
10\%, which is highly favorable.  Furthermore, adoption is far from
independently distributed, as the conditional probability that an app
user's contact joins the system is likely higher than that for a random
individual. So, even for lower values of $d$, the same critical threshold
may be sufficient due to positive correlation amidst the heterogeneity of
adoption.

As soon as that level of connectivity is reached, the amplification power
of our approach is unlocked. Each positive signal which is entered into the
system notifies thousands of people. Conversely, each user is able to
receive signals from thousands of other people, which facilitates user
acquisition and retention. The remaining subsections can then operate with
higher adoption thresholds.

\subsection{Accuracy of signals}

Next, we turn to analyze the potential discrepancy between the reports of
network distances to new signals based on the network of app users, and
actual network distances to those signals if they were calculated against
the full network of people (whether they are app users or not). One failure
mode corresponds to the situation where the actual network distance to a
new positive case is finite, but the user is at infinite distance according
to the network of app users (not connected, possibly because the positive
case is a non-user, or because of a non-user along the path). Another
failure mode corresponds to the app-reported distance being significantly
longer than the actual network distance, again due to non-users along the
shortest paths. The root cause therefore comes from two categories:
non-users along the path, and non-users who turn positive. The presence of
these sources of error make it imperative that in any deployment of our
approach, careful communication must be sent to inform the community that
the app is a tool to help increase caution, but a null signal should not
provide a false sense of security. Regardless, it is still a useful tool,
analogous to how convex passenger-side automobile mirrors are useful even
though they have blind spots and bear the legend ``objects in mirror are
closer than they appear.''

For non-users along the path, this becomes an interesting random network
structure question in itself: what is the effect on network distance of
missing nodes? We point to several influential factors which work to our
advantage, and which theoretically indicate that our approach would likely
yield favorable results. First, the typical structure of the physical
interaction network has ``small-world'' character, in that neighbors of a
given node tend to be more likely to be neighbors of each other. As
mentioned above, in the OpenABM-Covid19 model \cite{bib:openabm}, the
network is composed of three parts: a recurring household network, a
recurring occupation network, and some random interactions that change
daily. Digging deeper, their model of the household network is such that
each household has all pairs of interactions occurring every day, and each
occupation network is a Watts-Strogatz small-world network with a fixed set
of connections between individuals, where each day a random subset of half
of these connections is chosen as the interactions between individuals.
They perform additional refinements to account for networks of children.

In this type of model, a missing node $v$ does introduce significant
distortion in reported network distances between other nodes in the same
household as $v$ and other nodes in the same occupation network as $v$,
because $v$ was the main point of connection between those two worlds.
However, due to the small-world nature of the occupation network, that
sub-network is significantly more resilient to $v$'s absence, because there
are many alternative paths to consider that go around $v$. This indicates
that within networks corresponding to schools, universities, or large
workplaces, the network distance should be somewhat robust to missing
nodes, making our approach useful there even before widespread general
adoption.  Therefore, the most fragile paths are those of the form $uvwx$
between nodes $u$ and $x$ from different households, where the connection
$uv$ is between two nodes in one occupation network, the connection $vw$ is
between two nodes in another household network, and the connection $wx$ is
between two nodes in yet another occupation network. If either $v$ or $w$
is not participating, then the app-reported distance between $u$ and $x$
could be too high if there are no other short paths between $u$ and $x$.

Fortunately, the combination of these observations points to an example
pathway to impact: a city's high schools could adopt our approach as part
of its safe reopening plan, with an emphasis on all members of each
household joining the system (not only the student). Since many occupation
networks will have multiple connections to the high school network via
households, this will effectively leverage the structures of both the high
school network and the small-world nature of occupation networks to provide
the redundancy required to make the system resilient to non-participation.

The issue of positive cases occurring among non-users is easier to resolve.
The nature of infection spread is such that positive cases appear in
contiguous clusters when viewed in the context of the actual interaction
network. Consequently, even if only some fraction of the positive cases
occurs among app users, those positive case reports will still appear at
roughly the correct distances from every other user. Since the success of
our approach is not based on identifying and quarantining every exposure,
but rather providing situational awareness that inspires self-protective
caution, it is enough for us as long as the fraction of signal that appears
corresponds to nonzero actual signal appearing around the appropriate
distance.

\subsection{Impact of signal}

The previous subsections provided heuristic analysis of how partial
adoption affects the generation of signals of positive infection at various
network distances. Finally, we discuss the impact of these signals once
received. Our impact mechanism is based on engaging the natural
self-preservation instincts of people within short network distance of
positive cases. The nature of the communication does not emphasize
quarantine, but rather offers an opportunity to protect oneself from
others. It is not yet definitively known what the precise impact of various
precautions is, but an early research study on coronaviruses by the World
Health Organization \cite{bib:who-avoid} provided some figures which we may
use to estimate the order of magnitude of impact. Specifically, they were
moderately confident that ``a physical distance of more than 1 m probably
results in a large reduction in virus infection; for every 1 m further away
in distancing, the relative effect might increase 2.02 times.'' There has
been quantitative uncertainty around the effect of mask use, with
``relative risk (RR) reductions for infection ranging from 6--80\%'' stated
by Sch\"unemann et al. in \emph{The Lancet Respiratory Medicine}
\cite{bib:who-masks}, while a recent study by Asadi et al. in
\emph{Scientific Reports} \cite{bib:masks-davis} found that in terms of
particle emission, surgical masks blocked 90\% of particles while speaking,
while cloth masks increased the number of particles. That said, if people
choose to reduce their duration or frequency of in-person interaction, that
certainly reduces their risk by a corresponding rate.

Even though these quantitative figures are not definite, it appears likely
that if people in the vicinity of the cluster actively seek to protect
themselves from infection, they can reduce the infection transmission
probability by a factor that is large relative to $R_0$. And, as further
research emerges regarding self-protection techniques, people will be able
to adopt the most effective methods, whatever they may be. Furthermore, the
network distance information transcends the app userbase, because if a
household member is an app user who sees positive case(s) approaching, they
have a likelihood of informally alerting others in their household, even if
others are not app users. This information is still relevant and useful,
because if the original app user had a case at network distance $d$, then
the case is at distance at most $d+1$ from their household member. The same
phenomenon holds within occupational networks.  Therefore, in terms of the
impact delivered by our system, once there is a signal of nearby
positivity, its impact is likely to affect the behavior of not only the app
user, but also non-app users in the vicinity.  This represents a
fundamental distinction from the situation with anonymous post-exposure
notification apps. Indeed, the primary mechanism for impact of anonymous
post-exposure notification is via quarantine, but the anonymity inherently
makes it difficult to know which non-app users should quarantine, and it
would be highly inefficient for everyone nearby to quarantine as well, not
to mention the stigma of asking others to quarantine based on one's own
exposure.

In order to model the impact of a signal delivered by our approach, one
would need to estimate the probability $p_1$ that the recipient of a signal
of definitive nearby infection takes action to avoid infection, and the
probability $p_2$ that their action, if taken, interrupts a would-be
transmission. It is an added bonus that there is also some probability
$p_3$ that they tell people nearby. Since the purpose of this article is to
introduce our alternative approach, we leave it to future behavioral
science research to precisely estimate the actual values of these
parameters. We instead reason about their order of magnitude relative to
the existing post-exposure approach from a theoretical perspective.
Because of its alignment with self-protection, our first probability $p_1$
is likely to be an order of magnitude higher than the probability of a user
complying with an anonymous multi-day quarantine request, especially when
quarantine requests are sent to people whose current infection probability
is 10\% or below.  And, with proper communication and education, our $p_1$
can be increased because it is often directly in the user's self-interest
to protect themself; on the other hand, even with more education, it will
be extremely difficult to convince people to substantially inconvenience
themselves at low infection risk, unless there are no alternatives. It is
likely that our $p_2$ achieved by actively increasing vigilance is
comparable in order of magnitude to that from the post-exposure system
because although quarantine stops further transmission, post-exposure
notification systems lack the lead time of our approach, and due to delays
in testing, they may start the quarantine too late.  Finally, because there
is relatively little stigma attached to communicating the approach of
positive cases from afar (as compared to disclosing an actual exposure),
our $p_3$ is also substantial. Note, however, that the corresponding $p_3$
for post-exposure notification (inspiring multi-day voluntary quarantines
for nearby non-users of a post-exposure notification app after receiving a
low-infection-probability exposure notification) would be near-zero, and so
any reasonable value here already represents a significant advance. In
conclusion, the net impact of our signal is likely to be at a higher order
of magnitude than that of the signals sent by existing approaches. It still
is sensible to apply all approaches, however, because they control the
infection spread in different ways.

\section{Privacy and security}
\label{sec:privacy-security}

Many articles on digital interventions in the context of contact tracing
have discussed issues of privacy and security at great length
\cite{bib:ahmed-survey, bib:cambridge-contact-tracing, bib:lancet-privacy,
  bib:pact-washington, bib:cho-ippolito-yu, bib:harvard-privacy, bib:pact,
bib:sun-survey, bib:dp3t}. This section is written in a somewhat more
technical form, so as to more accurately address such issues. In the
context of the existing literature, our approach can be summarized as a
pseudonymous, centralized system which directly collects no personally
identifiable information and no GPS location information, but does collect
timestamped relative proximity information and pseudonymized Wi-Fi
information. As the purpose of this article is to introduce our approach of
informing individuals of their network distance from positive reports, as
opposed to detailing a specific app implementation, we have written this
section so that it serves as a preliminary privacy analysis of our general
concept. We do not claim to enumerate the entire attack space here, nor to
provide proofs of security.

Significant concerns about centralized systems revolve around what could be
done with access to the central database. An important initial observation
is that our central database does not strictly need to collect any more
information than Singapore's TraceTogether system
\cite{bib:trace-together}, and therefore our system represents a very
significantly more effective intervention for a comparable privacy loss to
the central authority, as compared to TraceTogether or any similar
centralized system. Indeed, even though we have an additional Wi-Fi sensing
layer which could be abused by a dishonest central authority to deduce
location data, any such centralized Bluetooth-only system could similarly
be compromised by positioning identified Bluetooth beacons in known
locations.  Alternatively, since we only use Wi-Fi to affirm relative
proximity, if one believes it is possible to have non-colluding entities,
it is possible to use a different implementation of the Wi-Fi system in
Section \ref{sec:network-construct} as follows. A completely separate and
non-colluding Wi-Fi Matching Entity could take the responsibility for
determining Wi-Fi matches, whereby each mobile app sends a single-use
random identifier representing its identity together with its Wi-Fi
information to the separate Wi-Fi Matching Entity, while informing the main
central server of which single-use random identifier it just sent.  Then,
the Wi-Fi Matching Entity informs the main central server of which pairs of
single-use random identifiers correspond to users who were proximal,
without disclosing any actual Wi-Fi information, and it permanently
destroys all of its data within hours of collection. The main central
server then looks up which real user UUID's correspond to the proximal
single-use random identifiers.  Importantly, the main central server finds
out which devices were proximal, without ever receiving any Wi-Fi
information, and the Wi-Fi Matching Entity is unable to deduce any user
identities from the single-use random identifiers. (It is worth noting that
although our approach avoids the use of GPS coordinates due to the general
perception of GPS as being privacy-invasive, a similar non-colluding-entity
technique could even be applied to anonymously determine relative proximity
relationships using GPS coordinates or other absolute location information
instead of Wi-Fi information, so that neither entity is independently able
to associate absolute location information with users.)

Due to our system's centralized operation, individual devices process very
little information: only the temporary random identifiers of the other
devices they were directly near, with times and relative distances.
Consequently, even if an individual device were hacked, it would yield only
one node's local information in the network. Centralization also prevents
individual devices from knowing the exact time that they encountered a
positive case (an issue for some decentralized approaches), because the
central server provides intentionally ambiguous ranges instead of precise
values. In particular, since our approach only seeks to trigger increased
caution, we do not continue sending proximity-based warnings even if a
positively tagged device continues to roam in public during its contagious
period.  It is sufficient for our purposes to generate signals only at the
time of reporting, as detailed in Section \ref{sec:chart-construct}. That
said, if a user was only ever around one other person, and they receive a
positive signal at network distance 1, then they have certainly learned
about the status of that other individual.  This type of edge case occurs
in many contact tracing methods, both digital or manual, and should be
handled via disclosure in the app's privacy policy.

\subsection{Abuse by the central authority itself}

It is true that if it were operating the central database, a government
could understand the interaction history between Person $A$ and Person $B$
by sending Agents $A'$ and $B'$ to follow each of Person $A$ and Person $B$
around, and then using the central database's records to start from the
timestamped interaction record of Agent $A'$, identify which UUID
corresponds to Person $A$, and then perform the same deduction from Agent
$B'$ to Person $B$, after which they could consult the database with the
now-known UUID's of Person $A$ and Person $B$. Therefore, it is beneficial
for our approach to be operated by a non-governmental entity. It is worth
mentioning that long before contact tracing apps, mobile telecom operators
already could deduce information about people's whereabouts due to cell
tower triangulation, and mobile operating system providers already could
deduce even more precise location information through their multi-sensor
location services. Indeed, the Israeli government acquired location
information from mobile telecom operators to control the spread of COVID-19
\cite{bib:israel}. Other concerns arise if the organization operating our
framework also has control over the integration between our framework and
the hardware (e.g., if the framework is integrated with the underlying
operating system of the phone), as it is in theory possible for personal
information entered via the operating system to mix with data collected by
our approach.

More computationally intensive attacks might be possible with sufficient
resources, with access to the full interaction network. For example, by
analyzing clustering within the interaction network, sub-networks
corresponding to major cities can be guessed. After sorting by population
and taking relative geography into account, the sub-networks could be
guessed to correspond to specific cities. Next, within a city's
sub-network, the further sub-networks containing a particularly high
density of frequently changing 2 a.m.\ interactions could be identified,
and those might correspond to universities. Again by comparing sizes of
those sub-networks, it might be possible to guess which universities
corresponded to each sub-network. Using the relative proximity distances,
it might be possible to identify further sub-networks corresponding to
individual dormitories, and then individual floors of dormitories. Finally,
it is conceivable that in some special cases if a student lives at the very
end of a hallway, the interaction network might identify them as an extreme
point. (However, the accuracy of the Bluetooth distance measurements would
likely be insufficient for that deduction, and ultrasonic distance
measurements do not penetrate walls.)

Although all of these attacks are theoretically possible, it is important
to note that many commonly used apps collect far more personal information.
For example, in theory one could apply facial recognition to the videos
uploaded on a social app to deduce a great deal of proximity information,
capturing much more additional information as well. Often, those other apps
are also explicitly monetizing that personal information by selling
advertisements. Yet many people flock to those apps because they perceive
direct value in the apps.  In the case of our approach, we seek to
similarly deliver a direct value proposition, while actively taking
measures to limit invasions of privacy.

\subsection{Attacks by users outside the central authority}

We consider two general classes of attacks by ordinary users who do not
have central database access: those who seek to compromise privacy by
deducing the network structure, and those which seek to disrupt civil
society by injecting false information. It is worth noting that the first
class of attacks can be conducted by ``semi-honest'' participants, who do
not actively disrupt the system's behavior, but record the information they
are given, and make more sophisticated deductions based upon it.

We begin by considering attacks aimed at disrupting society, e.g., by
sowing unfounded panic. Fortunately, the Bluetooth and ultrasound portions
of our protocol involve both parties in each interaction, and so it is
difficult for an attacker to unilaterally insert an interaction between
themselves and a legitimate user. They can, however, insert an interaction
between themselves and other illegitimate users that they control. However,
in order to affect legitimate users, they do still need to invest the
effort to physically come close to legitimate users, which is impractical
at large scale. Our Wi-Fi interactions can in theory be inserted
unilaterally and illegitimately (without actually being in proximity of the
Wi-Fi access point) if the app is decompiled and the Wi-Fi interaction
recording procedure is reverse-engineered. Fortunately, due to the fact
that the devices should still be constantly communicating via Bluetooth and
ultrasound, the structure of the network can be used to identify suspicious
devices which are inserting illegitimate Wi-Fi interactions.

In order to maliciously insert positive signals into any proper deployment
which uses positive test confirmation tokens as in Section
\ref{sec:positive-token}, the attacker needs to acquire valid one-time-use
tokens. This would involve either compromising a token-issuing authority,
or acquiring tokens that were distributed to others and entering them into
the wrong app. These attacks can be mitigated by expiring tokens shortly
after their issuance, and by limiting a token's usage scope (e.g., only to
a certain community of users who has opted into working with a specific
token-issuing authority).

We finish by considering attacks which seek to deduce the network
structure. By sending only transient identifiers during Bluetooth
communication, our system is more robust to ``linkage'' attacks which seek
to identify the same user over time based on their Bluetooth transmission
activity. On the other hand, by providing users with charts of the number
of positive cases (and the numbers of individuals) at various network
distances from them, our system does indeed supply additional information
that could be used to deduce some elements of network structure.

For example, in the following situation it is possible to deduce that
Person $A$ and Person $B$ did not spend substantial time together in the
last 14 days. Agent $A'$ is deployed to spend at least 15 minutes around
Person $A$, while Agent $B'$ is deployed to do the same with Person $B$.
Then, Agent $A'$ reports a positive case. If Agent $B'$ does not see a
positive case appear in their chart at distance less than or equal to 3,
then Person $A$ and Person $B$ did not spend time with each other over the
last 14 days.

The converse does not immediately hold. Even if Agent $B'$ sees a positive
case appear at distance 3, that is not definitely because Person $A$ and
Person $B$ spent time together, because it could be that when Agent $A'$
built a connection to Person $A$, Agent $A'$ also accidentally built a
connection to Person $C$ who happened to be within 10 meters at the time,
and Person $C$ was the one who had spent time with Person $B$. Or, it could
be that an unrelated positive case was reported at around the same time,
and showed up at distance 3 from Agent $B'$. These challenges can be
overcome with a more complicated attack in which the agents ensure that
there is nobody else within 10 meters of them besides their target, and
send multiple positive reports to increase the confidence that Person $A$
and Person $B$ had spent time together. Note, however that this still does
not indicate what time within the 14-day window such an interaction could
have occurred, or for exactly how long (beyond 15 minutes). And, if Person
$A$ and Person $B$ meet again within 14 days of their last meeting, it will
be impossible to periodically repeat the same style of attack to deduce the
earlier meeting time, because even the 14-day window immediately following
that earlier meeting time will contain an interaction that constitutes a
connection (the result will look the same).

In summary, it is true that some network information can be deduced even
without access to the central database, and we leave it to the reader to
judge whether access to information such as the above is a worthwhile
tradeoff for a system which can provide early warning of an approaching
infection.

\section{Conclusion}

In this article, we have introduced a novel approach which empowers every
individual member of society to actively avoid infection. It uses the
contact tracing network ``in reverse,'' where instead of asking everyone
within distance 1 of a positive case to quarantine, it tells everyone how
far away the new positive cases have struck in their physical interaction
network. This reversal changes the nature of the intervention, from one
which ``protects others from you'' to one which ``protects you from
others.'' Through that flip, the incentive structure also reverses, as
users are given the opportunity to protect themselves before it is too
late. Suddenly, users prefer false positives over false negatives (``better
safe than sorry''), which is the opposite of the situation when they use
apps that ask them to quarantine (the culturally unfamiliar ``guilty until
proven innocent''). The final reversal is what enables us to use Wi-Fi as a
sensor, as Wi-Fi access point overlap generates too many false positives
for post-exposure notifications, yet extended duration on the same Wi-Fi
access point fairly accurately captures much of the underlying physical
interaction network, which is all we need. As discussed in our heuristic
analysis, the potential impact of our approach is very significant, likely
reaches critical mass at low adoption rates, and even is likely to spill
over to alert non-users. It is unrealistic to expect there to be a perfect
solution, but our theoretical heuristics indicate that on balance, this
could be a highly beneficial intervention. 

It is natural to ask why such a technique was never used before. The answer
lies in the current state of technology. It is only in the past decade that
the necessary transmitters commonly appeared on smartphones, and those
smartphones proliferated. Wi-Fi has certainly been around for some time,
but our other Bluetooth and ultrasound sensors augment the system in
important ways. Conveniently, in the midst of this devastating COVID-19
pandemic, we just happen to be in a position where we can deploy this new
system at global scale.

That said, the author acknowledges that his primary expertise is in
mathematics and technology, and the problem of pandemic control is complex,
requiring expertise in other disciplines. Therefore, the purpose of this
article is not to claim a proof of a solution to the pandemic. On the
contrary, the reason for putting much of the logic of this article into
words was to produce a written document which could invite comment,
discussion, and collaboration. This article opens more questions than it
answers, acknowledging that in order to definitively understand the
potential of this approach, there are many questions in public health,
medicine, biology, modeling, behavioral science, and security that need to
be answered. If practice aligns sufficiently with theory, then this new
approach could equip everyone with the equivalent of a personal weather
satellite, to protect themselves by tracking an otherwise-invisible disease
approaching from afar.

\section*{Acknowledgments}

Thanks to Anna Bershteyn, Timothy Chu, Pete Hoch, Ian McCullough, Janet
Mertz, Francesmary Modugno, Philip Welkhoff, Lowell Wood, Shannon Yee, and
Yun William Yu for providing valuable feedback on earlier versions of this
manuscript.


\begin{thebibliography}{99}
  \bibitem{bib:openabm-google} Abueg, M., Hinch, R., Wu, N., Liu, L.,
    Probert, W., Wu, A., Eastham, P., Shafi, Y., Rosencrantz, M., Dikovsky,
    M., Cheng, Z., Nurtay, A., Abeler-D\"orner, L., Bonsall, D., McConnell,
    M., O'Banion, S., and Fraser, C. (2020). \emph{Modeling the combined
    effect of digital exposure notification and non-pharmaceutical
    interventions on the COVID-19 epidemic in Washington state}. MedRxiv.
    \url{https://doi.org/10.1101/2020.08.29.20184135}

  \bibitem{bib:ahmed-survey} Ahmed, N., Michelin, R., Xue, W., Ruj, S.,
    Malaney, R., Kanhere, S., Seneviratne, A., Hu, W., Janicke, H., and
    Jha, S. (2020, July 27). \emph{A Survey of COVID-19 Contact Tracing Apps
    }. ArXiv. \url{https://arxiv.org/abs/2006.10306}

  \bibitem{bib:israel} Altshuler, T., and Hershkowitz, R. (2020, July 6).
    \emph{How Israel's COVID-19 mass surveillance operation works.}
    The Brookings Institution Tech Stream.
    \url{https://www.brookings.edu/techstream/how-israels-covid-19-mass-surveillance-operation-works/}

  \bibitem{bib:gaen-apple} Apple, Inc. (2020, September 26).
    \emph{Privacy-Preserving Contact Tracing}.
    \url{https://covid19.apple.com/contacttracing}

  \bibitem{bib:masks-davis} Asadi, S., Cappa, C., Barreda, S., Wexler, A.,
    Bouvier, N., and Ristenpart, W. (2020). Efficacy of masks and face
    coverings in controlling outward aerosol particle emission from
    expiratory activities. \emph{Scientific Reports} 10, 15665.
    \url{https://doi.org/10.1038/s41598-020-72798-7}

  \bibitem{bib:trace-together} Baharudin, H. (2020, March 20).
    Coronavirus: Singapore develops smartphone app for efficient contact
    tracing. \emph{The Straits Times}.
    \url{https://www.straitstimes.com/singapore/coronavirus-singapore-develops-smartphone-app-for-efficient-contact-tracing}

  \bibitem{bib:behavior-survey-nature} van Bavel, J., Baicker, K., Boggio,
    P., Capraro, V., Cichocka, A., Cikara, M., Crockett, M., Crum, A.,
    Douglas, K., Druckman, J., Drury, J., Dube, O., Ellemers, N., Finkel,
    E., Fowler, J., Gelfand, M., Han, S., Haslam, S., Jetten, J., Kitayama,
    S., Mobbs, D., Napper, L., Packer, D., Pennycook, G., Peters, E.,
    Petty, R., Rand, D., Reicher, S., Schnall, S., Shariff, A., Skitka, L.,
    Smith, S., Sunstein, C., Tabri, N., Tucker, J., Van der Linden, S., Van
    Lange, P., Weeden, K, Wohl, M., Zaki, J., Zion, S., and Willer, R.
    (2020). Using social and behavioural science to support COVID-19
    pandemic response. \emph{Nature Human Behavior}, 4(5), 460--471.
    \url{https://doi.org/10.1038/s41562-020-0884-z}

  \bibitem{bib:cambridge-contact-tracing} Bell, J., Butler, D., Hicks, C.,
    and Crowcroft, J. \emph{TraceSecure: Towards Privacy Preserving Contact
    Tracing}. ArXiv. \url{https://arxiv.org/abs/2004.04059}

  \bibitem{bib:lancet-privacy} Bengio, Y., Janda, R., Yu, Y., Ippolito, D.,
    Jarvie, M., Pilat, D., Struck, B., Krastev, S., and Sharma, A. (2020).
    The need for privacy with public digital contact tracing during the
    COVID-19 pandemic. \emph{The Lancet Digital Health}, 2(7), E342--E344.
    \url{https://doi.org/10.1016/S2589-7500(20)30133-3}

  \bibitem{bib:forbes-klick} Bowen, D. (2020, July 22). These 3 tools could help
    assess your risk of getting COVID-19. \emph{Fast Company}.
    \url{https://www.fastcompany.com/90530405/these-tech-tools-could-help-mitigate-covid-19-risks}

  \bibitem{bib:nhs-qr} Burgess, M. (2020, September 24). Everything you
    need to know about the new NHS contact tracing app. \emph{Wired
    Magazine}.
    \url{https://www.wired.co.uk/article/nhs-covid-19-tracking-app-contact-tracing}

  \bibitem{bib:pact-washington} Chan, J., Foster, D., Gollakota, S.,
    Horvitz, E., Jaeger, J., Kakade, S., Kohno, T., Langford, J., Larson,
    J., Sharma, P., Singanamalla, S., Sunshine, J., and Tessaro, S. (2020,
    May 7). \emph{PACT: Privacy Sensitive Protocols and Mechanisms for Mobile
    Contact Tracing}. ArXiv. \url{https://arxiv.org/abs/2004.03544}

  \bibitem{bib:taiwan} Cheng, H.-Y., Jian, S.-W., Liu, D.-P., Ng, T.-C.,
    Huang, W.-T., and Lin, H.-H. (2020). Contact Tracing Assessment of
    COVID-19 Transmission Dynamics in Taiwan and Risk at Different Exposure
    Periods Before and After Symptom Onset. \emph{JAMA Intern Med.},
    180(9), 1156--1163.
    \url{https://doi.org/10.1001/jamainternmed.2020.2020}

  \bibitem{bib:cho-ippolito-yu} Cho, H., Ippolito, D., and Yu, Y. (2020,
    March 20).  \emph{Contact Tracing Mobile Apps for COVID-19: Privacy
    Considerations and Related Trade-offs}. ArXiv.
    \url{https://arxiv.org/abs/2003.11511}

  \bibitem{bib:who-avoid} Chu, D., Akl, E., Duda, S., Solo, K., Yaacoub,
    S., and Sch\"unemann, H. (2020). Physical distancing, face masks, and
    eye protection to prevent person-to-person transmission of SARS-CoV-2
    and COVID-19: a systematic review and meta-analysis. \emph{The Lancet},
    395(10242), 1973--1987.
    \url{https://doi.org/10.1016/S0140-6736(20)31142-9}

  \bibitem{bib:tracefi} Cobb, S. (2020, August 2). Harvard to Track
    Affiliates' Wi-Fi Signals as Part of Contact Tracing Pilot. \emph{The
    Harvard Crimson}.
    \url{https://www.thecrimson.com/article/2020/8/2/tracefi-wifi-contract-tracing-coronavirus/}

  \bibitem{bib:wiki-covid-19-apps} COVID-19 apps. (2020, September 26). In
    \emph{Wikipedia}.
    \url{https://en.wikipedia.org/w/index.php?title=COVID-19_apps&oldid=980316587}

  \bibitem{bib:centralized-vs-decentralized} Criddle, C. and Kelion, L.
    (2020, May 7). Coronavirus contact-tracing: World split between two
    types of app. \emph{BBC News}.
    \url{https://www.bbc.com/news/technology-52355028}

  \bibitem{bib:reddit-gatech} GandalfBlue12. (2020, August 23). Your
    Reminder on NOVID. \emph{Reddit r/gatech.}
    \url{https://www.reddit.com/r/gatech/comments/if3sw2/your_reminder_on_novid/}

  \bibitem{bib:uga-gatech} GaNun, J. (2020, September 17). UGA, Georgia
    Tech students discuss contact tracing app NOVID in use at Tech.
    \emph{The Red \& Black}.
    \url{https://www.redandblack.com/uganews/uga-georgia-tech-students-discuss-contact-tracing-app-novid-in-use-at-tech/article_db5dd628-ef23-11ea-aa7c-238f25b13c38.html}

  \bibitem{bib:france-germany-uk} Ghosh, S. (2020, April 25). Apple is
    locked in a power battle with the UK, France, and Germany about how
    COVID-19 should be tracked. \emph{Business Insider}.
    \url{https://www.businessinsider.com/france-uk-apple-contact-tracing-apps-2020-4}

  \bibitem{bib:gaen-google} Google, Inc. (2020, September 26).
    \emph{Exposure Notifications: Using technology to help public health
    authorities fight COVID‑19}.
    \url{https://www.google.com/covid19/exposurenotifications/}

  \bibitem{bib:harvard-privacy} Hart, V., Siddarth, D., Cantrell, B.,
    Tretikov, L., Eckersley, P., Langford, J., Leibrand, S., Kakade, S.,
    Latta, S., Lewis, D., Tessaro, S., and Weyl, G. (2020, April 3).
    Outpacing the Virus: Digital Response to Containing the Spread of
    COVID-19 while Mitigating Privacy Risks. \emph{Edmond J. Safra Center
    for Ethics COVID-19 Rapid Response Impact Initiative White Paper 5}.
    \url{https://ethics.harvard.edu/outpacing-virus}

  \bibitem{bib:openabm} Hinch, R., Probert, W., Nurtay, A., Kendall, M.,
    Wymant, C., Hall, M., Lythgoe, K., Cruz, A., Zhao, L., Stewart, A.,
    Ferretti, L., Montero, D., Warren, J., Mather, N., Abueg, M., Wu, N.,
    Finkelstein, A., Bonsall, D., Abeler-D\"orner, L, and Fraser, C.
    (2020).
    \emph{OpenABM-Covid19 --- an agent-based model for non-pharmaceutical
    interventions against COVID-19 including contact tracing}. MedRxiv.
    \url{https://doi.org/10.1101/2020.09.16.20195925}

  \bibitem{bib:google-maps} Jee, C. (2020, September 24). Google Maps now
    shows you where covid-19 cases are spiking. \emph{MIT Technology
    Review}.
    \url{https://www.technologyreview.com/2020/09/24/1008865/google-maps-now-shows-you-where-covid-19-cases-are-spiking/}

  \bibitem{bib:book-kahn} Kahn, J. (Ed.). (2020). 
    \emph{Digital Contact Tracing for Pandemic Response: Ethics and
    Governance Guidance}. Baltimore: Johns Hopkins University Press.
    \url{doi:10.1353/book.75831}

  \bibitem{bib:tcd-bluetooth} Leith, D. and Farrell, S. (2020, October).
    Coronavirus Contact Tracing: Evaluating The Potential Of Using
    Bluetooth Received Signal Strength For Proximity Detection. \emph{ACM
    Computer Communications Review}.

  \bibitem{bib:tcd-bus} Leith, D. and Farrell, S. (2020, June 15).
    \emph{Measurement-Based Evaluation Of Google/Apple Exposure
    Notification API For Proximity Detection In A Commuter Bus}. ArXiv.
    \url{https://arxiv.org/abs/2006.08543}

  \bibitem{bib:tcd-tram} Leith, D. and Farrell, S. (2020, accepted for
    publication). Measurement-Based Evaluation Of Google/Apple Exposure
    Notification API For Proximity Detection In A Light-Rail Tram.
    \emph{PLOS ONE}. \url{https://www.scss.tcd.ie/Doug.Leith/pubs/luas.pdf}

  \bibitem{bib:loh-novid-accuracy} Loh, P. (2020). Accuracy of
    Bluetooth-Ultrasound Contact Tracing: Experimental Results from NOVID
    iOS Version 2.1 Using Five-Year-Old Phones. \emph{NOVID Technical
    Report}. \url{https://www.novid.org/downloads/20200626-accuracy.pdf}

  \bibitem{bib:behavior-survey-jbpa} Lunn, P., Belton, C., Lavin, C.,
    McGowan, F., Timmons, S., and Robertson, D. (2020). Using Behavioral
    Science to help fight the Coronavirus. \emph{Journal of Behavioral
    Public Administration}, 3(1).
    \url{https://doi.org/10.30636/jbpa.31.147}

  \bibitem{bib:novid-app} Maderer, J. (2020, September 16).
    \emph{CMU-Created COVID Early Warning App to Assist Campus Community}.
    Carnegie Mellon University News.
    \url{https://www.cmu.edu/news/stories/archives/2020/september/novid-to-cmu.html}
    
  \bibitem{bib:techcrunch-15-apps} Moss, A., Spelliscy, C., and Borthwick,
    J. (2020, June 5). \emph{TechCrunch}.
    \url{https://techcrunch.com/2020/06/05/demonstrating-15-contact-tracing-and-other-tools-built-to-mitigate-the-impact-of-covid-19/}

  \bibitem{bib:cmu-novid-accuracy} Payne, E. (2020, June 30).
    \emph{NOVID Is the Most Accurate App for Contact Tracing}. Carnegie
    Mellon University News.
    \url{https://www.cmu.edu/news/stories/archives/2020/june/novid-update.html}

  \bibitem{bib:sk} Park, Y., Choe, Y., Park, O., Park, S., Kim, Y.-M., Kim,
    J., Kweon, S., Woo, Y., Gwack, J., Kim, S., Lee, J., Hyun, J., Ryu, B.,
    Jang, Y., Kim, H., Shin, S., Yi, S., Lee, S., Kim, H., Lee, H., Jin,
    Y., Park, E., Choi, S., Kim, M., Song, J., Choi, S., Kim, D., Jeon, B.,
    Yoo, H., and Jeong, E. (2020). Contact Tracing during Coronavirus
    Disease Outbreak, South Korea, 2020. \emph{Emerging Infectious
    Diseases}, 26(10), 2465--2468.
    \url{https://dx.doi.org/10.3201/eid2610.201315}

  \bibitem{bib:pact} Rivest, R., Callas, J., Canetti, R., Esvelt, K.,
    Gilmor, D., Kalai, Y., Lysyanskaya, A., Norige, A., Raskar, R., Shamir,
    A., Shen, E., Soibelman, I., Specter, M., Teague, V., Trachtenberg, A.,
    Varia, M., Viera, M., Weitzner, D., Wilkinson, J., and Zissman, M.
    (2020, April 8). The PACT protocol specification version 0.1.
    \url{https://pact.mit.edu/wp-content/uploads/2020/04/The-PACT-protocol-specification-ver-0.1.pdf}

  \bibitem{bib:germany-gaen-disappoint} Rosenbach, M. and Schmergal, C.
    (2020, September 24). Germans Disappointed by Coronavirus Tracking
    App. \emph{Der Spiegel}.
    \url{https://www.spiegel.de/international/germany/lots-of-work-but-little-utility-germans-disappointed-by-coronavirus-tracking-app-a-7c30191e-b225-4c37-917d-41dc2a6078a1}

  \bibitem{bib:isle-wight} Sabbagh, D. and Hern, A. (2020, June 18). UK
    abandons contact-tracing app for Apple and Google model.  \emph{The
    Guardian}.
    \url{https://www.theguardian.com/world/2020/jun/18/uk-poised-to-abandon-coronavirus-app-in-favour-of-apple-and-google-models}

  \bibitem{bib:who-masks} Sch\"unemann, H., Akl, E., Chou, R., Chu, D.,
    Loeb, M., Lotfi, T., Mustafa, R., Neumann, I., Saxinger, L., Sultan,
    S., and Mertz, D. (2020). Use of facemasks during the COVID-19
    pandemic. \emph{The Lancet Respiratory Medicine}.
    \url{https://doi.org/10.1016/S2213-2600(20)30352-0}

  \bibitem{bib:nature-privacy} Sharma, T., and Bashir, M. (2020). Use of
    apps in the COVID-19 response and the loss of privacy protection.
    \emph{Nature Medicine}, 26, 1165--1167.
    \url{https://doi.org/10.1038/s41591-020-0928-y}

  \bibitem{bib:usa-9-million} Smith, M., Romero, S., and Nieto del Rio, G.
    (2020, October 29). U.S. Coronavirus Cases Surpass 9 Million With No
    End in Sight. \emph{The New York Times}.
    \url{https://www.nytimes.com/2020/10/29/us/coronavirus-nine-million-cases.html}

  \bibitem{bib:marketwatch-gps} Sokolow, A. (2020, July 9). U.S.
    contact-tracing efforts on Covid-19 falter on privacy worries,
    technology gap. \emph{MarketWatch}.
    \url{https://www.marketwatch.com/story/us-contract-tracing-efforts-on-covid-19-falter-on-privacy-worries-technology-gap-2020-07-09}

  \bibitem{bib:sun-survey} Sun, R., Wang, W., Xue, M., Tyson, G., Camtepe,
    S., and Ranasinghe, D. (2020, July 22). \emph{Vetting Security and
    Privacy of Global COVID-19 Contact Tracing Applications}. ArXiv.
    \url{https://arxiv.org/abs/2006.10933}

  \bibitem{bib:gatech-deploy} Toon, J. (2020, August 17). \emph{NOVID
    Exposure Notification App Enlists Smartphones in Coronavirus Battle}.
    Georgia Tech News Center.
    \url{https://news.gatech.edu/2020/08/17/novid-exposure-notification-app-enlists-smartphones-coronavirus-battle}

  \bibitem{bib:dp3t} Troncoso, C., Payer, M., Hubaux, J.-P., Salath\'e, M.,
    Laurus, J., Bugnion, E., Lueks, W., Stadler, T., Pyrgelis, A.,
    Antonioli, D., Barman, L., Chatel, S., Paterson, K., \v{C}apkun, S.,
    Basin, D., Beutel, J., Jackson, D., Roeschlin, M., Leu, P., Preneel,
    B., Smart, N., Abidin, A., G\"urses, S., Veale, M., Cremers, C.,
    Backes, M., Tippenhauer, N., Binns, R., Cattuto, C., Barrat, A., Fiore,
    D., Barbosa, M., Oliveira, R., and Periera, J. (2020, May 25).
    \emph{Decentralized Privacy-Preserving Proximity Tracing}.
    \url{https://github.com/DP-3T/documents/blob/master/DP3T\%20White\%20Paper.pdf}

  \bibitem{bib:cdc-contact-tracing} United States Centers for Disease
    Control and Prevention. (2020, September 10). \emph{Contact Tracing for
    COVID-19}.
    \url{https://www.cdc.gov/coronavirus/2019-ncov/php/contact-tracing/contact-tracing-plan/contact-tracing.html}

  \bibitem{bib:covidwatch} Wilson, A., Aviles, N., Petrie, J., Beamer, P.,
    Szabo, Z., Xie, M., McIllece, J., Chen, Y., Son, Y.-J., Halai, S.,
    White, T., Ernst, K., and Masel, J. (2020, September 17).
    \emph{Quantifying SARS-CoV-2 infection risk within the Google/Apple
      exposure notification framework to inform quarantine
    recommendations}. MedRxiv.
    \url{https://doi.org/10.1101/2020.07.17.20156539}

  \bibitem{bib:wsj-apps-not-ready} Winkler, R. and Haggin, P. (2020, June
    22). America Is Reopening. Coronavirus Tracing Apps Aren't Ready.
    \emph{The Wall Street Journal}.
    \url{https://www.wsj.com/articles/america-is-reopening-coronavirus-tracking-apps-arent-ready-11592845646}
    
    % https://lasec.epfl.ch/people/vaudenay/swisscovid.html

\end{thebibliography}
\end{document}